# TYPES OF PARADOX IN PHYSICS

DRAGOLJUB CUCIĆ

*Regional centre for talents Mihajlo Pupin, Pancevo, Serbia, rctpupin@gmail.com*

ABSTRACT

Paradoxes are a relatively frequent occurrence in physics. The nature of their genesis is diverse and they are found in all branches of physics. There are a number of general and special classifications of paradoxes, but there are no classifications of paradoxes in physics. Nowadays, physics is a fundamental and rather formalized science, the paradoxes of which imply falsity and imprecision. One of the basic methods of addressing a problem is to present classifications that facilitate its formulation and study. This work groups together the paradoxes in physics according to certain common characteristics, which should assist in explaining the causes for paradox formation.

*There is a mistake somewhere...*

**Lewis Carroll**, *Alice's Adventures in Wonderland*

**Introduction**

Paradox, as a phenomenon, along with the examples that illustrate it, is the subject of many books and papers in mathematics, logic, and philosophy. These areas of human spirituality and intellect offer syntheses that attempt to generalize and classify paradoxes according to common factors.

Willard Quine (Willard Van Orman Quine, 1908-2000) in his book *The Ways of Paradox* (1966) is the first to make a seriously accepted systematization of paradoxes, classifying them in three categories. In addition to Quine it is necessary to mention Frank Ramsey (1903–1930) who divided paradoxes into two basic groups.[1] The first known paradoxes date back to ancient Greece. An excellent study of the paradoxes from that time can be found in Roy Sorensen's book *A Brief History of the Paradox*: *philosophy and the labyrinths of the mind* (2003). I would especially like to mention the book *Paradoxes* (1987) by Richard Mark Sainsbury as one of the most successfully written books on paradoxes.

In Serbia, among the books dealing with paradoxes, a distinguished place belongs to *Space, Time, Zeno* (1986) a book by Milos Arsenijevic, professor at Belgrade University, Faculty of Philosophy. However, this book deals with the *paradoxes of Zeno* from a narrowly professional point of view. Miomir Jaksic, professor of Belgrade University, Faculty of Economics, in his book *Paradoxes and Riddles of Economics* (1998) deals with paradoxes in economics. He enumerates and explains them, trying to make a classification synthesis of the given paradoxes. In doing this Jaksic distinguishes three types of paradox in economics: everyday vague observations, explained paradoxes and inexplicable paradoxes. Common denominators for paradoxes in economics are, according to Miomir Jaksic, "unproved assumptions" that they are based upon. Rozalia Madarasz-Szilagyi from the University of Novi Sad, in her electronic course book/hand book (2004) for her course *Mathematical logic and set theory* deals with paradoxes of sets in mathematics while not attempting to define the problem of paradox in mathematics. Analysing the trends of contemporary mathematics, appearance of set theory, as well as the basis of mathematics in the late 19th and early 20th century, *Hilbert's programme* and *Russell-Whitehead formalization*, Milan Bozic, professor of Belgrade University, in his book *Study of history and philosophy of mathematics* (2002) mentions paradox as a phenomenon in mathematics, but only in a historical sense.

---
[1] Footnote 11.

From all the courses on paradox that I have been able to access, I would like to mention the one by Laurence Goldstein, professor at the University of Kent in England. This course can be accessed on the internet.

There have been many workshops dealing with the role of paradox in cognitive sciences and its semantic role.

It could be noticed that physics and paradoxes in physics have not been mentioned above. I haven't been able to find any. There are a number of papers on individual paradoxes in physics that analyse paradoxes and offer their solutions. These individual paradoxes in physics are often very well known even to physics laymen.

At the *Wikipedia* site paradoxes in physics are enumerated, albeit in a very incomplete and erratic way. However, it can serve as a starting point for finding sources which in turn deal only with individual paradoxes in physics. No attempt is made to form any classification, compare to what has been done in mathematics, logic and philosophy, or to classify and synthesize opinions about paradoxes in physics.

It would be significant to mention Yakir Aharonov and Daniel Rohrlich's book *Quantum Paradoxes: Quantum Theory for the Perplexed* (2003) that enumerates and explains the paradoxes of quantum mechanics on a very high analytical level.

The aim of this work is to determine the epistemological significance of the existence and formation of paradoxes in physics and to clearly define their structure in order to make them more easily recognizable. Also, the intention is to make a classification of paradoxes in physics according to certain common classification factors, and to make paradox as a physical phenomenon a clear, recognizable and acceptable category from the viewpoint of philosophical and historical science. The attempt has been made to form a connection between a physicist's world image, based on observation and experiments, and that of a philosopher who tends to establish "*...conceptual scheme through a logical study of semantics.*" (Sorensen, 2003, p. 369)

Since the paradoxes in physics have been dealt with solely as individual phenomena, attempting no common formula, this work is going to do precisely that. By generalising a large number of examples of paradoxes in physics[2], an empirical conclusion about their classification is synthesized.

**Paradoxes**

Paradoxes are created when the proper "tools" are not used, when certain phenomena are interpreted using inappropriate terms, due to the accepted meanings of the terms being inadequate. This is what Bertrand Russell noticed when he pointed out the problems in discussion caused by failure to distinguish between word disputes and fact disputes[3], which is what defines the distinction between semantic and ontological perplexities that provide the basis of the paradox. The terms that mean one thing in one branch of a certain discipline, mathematics or physics for example, do not mean the same in another branch.

---

[2] What is considered to be a paradox in physics.
[3] Russell, B. (1925).

The basic mistake is the irrational desire to generalize a *principle of uniformity*, which makes it possible for the inconsistencies to appear and bring about the paradox.

Methods of reasoning are defined by rules. Different methods of reasoning "comply" with different rules. There is no universal method of reasoning that provides the correct conclusion in different situations. There is no sum of rules that would invariably lead the subject to the correct conclusion. Whether the established opinion and conclusion is arrived at by deductive or inductive means, by various syllogisms, sophistic reasoning, or by axiomatics based on rationalistic norms or empirics, it proves that there is no privileged mode of thought, and that each mode has its advantages and disadvantages. These advantages and disadvantages appear in different situations and demand, according to the situation, a special necessary way of reasoning as a currently valid form of thought. Kurt Gödel (1906-1978), in his 1931 book *On Formally undecidable Propositions of Principia Mathematica and related Systems* formulates a new model of mathematically logical thought. His two theorems prove that in every strictly formal mathematical system there are statements that can be neither confirmed nor refuted, although they are formed by axioms built in the system. In simpler terms this means that: ***axioms of a formal system allow for the possibility of contradiction***. This implies a real possibility of existence of paradox in any formalized system, in our case – physics. The theorems we are relying in these assumptions are:

- •If the axioms of a certain theory are not contradictory, there are theorems that can prove neither that they are valid nor that they are invalid.

- There is no precise way to prove that the axioms of a certain theory are not contradictory.

Theorems are cited here as a principle basis that enables the existence of causes for the appearance of paradoxes in physics.

The desire to formalize thought dates from the times of Ancient Greece[4] and Aristotle (384-322 BC), with the "proof of truth" of what was and what still is, being the unreachable "Holy Grail" of the search of knowledge.

In the lectures of Laurence Goldstein, posted on the internet, in which he deals with the phenomenon of paradox, he defined paradox as:

> "*A train of reasoning that leads from premises that seem obviously true, via apparently impeccable steps of reasoning, to a conclusion that is contradictory or crazy.*"

Goldstein noticed that paradoxes create the possibility for questions that lead to inconsistency and that to solve a paradox it is sometimes necessary to re-examine even the fundamental premises. Goldstein's lectures are very helpful in making a complete study of authors who work on the phenomenon of paradox. I would like to emphasize his observation that paradoxes should not be solved by the so-called "bad arguments". He explains his opinion by calling upon the works of different authors to illustrate *the Paradox of Epimenides' liar*. Goldstein also classifies paradoxes[5] according to "depth", the paradoxes that are deeper being more significant than the less deep ones. He analyzes various, well-known paradoxes, types of paradoxes and the opinions of author who write about paradoxes.

---

[4] in the traditions of European judeo-christian culture.
[5] Similar to Sainsbury.

Sainsbury deals with paradoxes in a more generalized, serious and complex way than many of the aforementioned authors. On the subject of paradox he writes:

> *"This is what I understand by a paradox: an apparently unacceptable conclusion derived by apparently acceptable reasoning from apparently acceptable premises. Appearances have to deceive, since the acceptable cannot lead by acceptable steps to the unacceptable. So, generally, we have a choice: either the conclusion is not really unacceptable, or else the starting point, or the reasoning, has some non-obvious flaw."*
> (Sainsbury, 1995.)

According to Sainsbury paradoxes are a form of amusement, but at the same time a very serious phenomenon. The increase in paradox frequency is evident in crises and revolutionary leaps in thinking. Sainsbury's opinion coincides with Kuhn's conclusion that the "instability" of systems of reasoning is clearly recognizable when the existing paradigm is in a crisis or in a more extreme case, **when it is changed**!

For Gareth Matthews paradox is a conflict with conceptual truth. For John Leslie Mackie (1917-1981) it is contained within the proof, which is a classical position of formal logic dedicated to the study of truthfulness of proof and proving as a set of regularities.

Roy Sorensen understands paradox as a form of a riddle. He also identifies parts of the riddle with paradox. He compares his claim with the statement: "*parts of a rose can be called a rose*". In doing so, he points out that it matters not whether the rose is a bush or a single flower. According to him, the reason why paradoxes often cannot be solved is the fact that the prerequisites leading to paradoxes do not include all the possibilities, which at the same time brings about the paradox.

> "*Those who develop the logic of questions define direct answer as an answer given with as much information as the questioner wanted.*" (Sorensen, 2003.)

I would like to point out the opinion of Nicholas Rescher who claims that:

> "*Paradox is the product not of a mistake in reasoning but of a defect of substance: a dissonance of endorsements*" (Rescher, 2001.)

Paradox is a multidisciplinary phenomenon. It appears in every field of human activity and it is necessary to find common denominators between the paradoxes that appear in these different fields, which can be connected into a general synthesis dealing with the explanation of paradox appearance.

One of the founders of "paradoxism"[6], Douglas R. Hofstadter, noticed that:

---

[6] An invented term denoting the inclination to observe and analyze phenomena by paradox.

> *"...the drive to eliminate paradoxes at any cost, especially when it requires the creation of highly artificial formalisms, puts too much stress on bland consistency, and too little on the quirky an bizarre..."* (Hofstadter, 1979, p. 22)

... which brings us again to a world completely alien to human understanding of reality. Marianne W. Lewis from the University of Cincinnati in the USA deals with the phenomenon of paradox in organization theories, research methods, technology and innovation management. She researches the phenomenon of paradox that disrupts and prevents innovation. She considers paradox to be a "tool" for defining problems and efficiency enhancement. Marianne Lewis describes paradox as a discrepancy the elements of which are interconnected in an apparently logical but simultaneously absurd way. According to her the purpose of paradoxes is not to be solved but to make a contribution to the search of truth and to the process of learning. She sees paradoxes as real objects that can be discovered. In her view paradox is:

1. a conflicting interpretation of an individual phenomenon,
2. a conflicting opinion,
3. aid in understanding diverging interpretations,
4. perceptive,
5. an appropriate illusion caused by social interaction,
6. identification of different views,
7. a characteristic of the observer – not of the observed,

8. a possible consequence used negatively to define something.

Mike Metcalfe, professor at the University of South Australia, and a leading authority on the field of strategy thinking using pragmatic systems, writes about Lewis' understanding of paradox:

> *"Paradoxes are used to find, explain and justify the existence of alternative interpretations which it is thought always exist around any understanding of a complex social phenomenon. Striving to find, remove or work with paradox is thought to be insufficient; rather, paradox needs to be seen as a window through which to creatively appreciate the world."* (Metcalfe, 2005)

Based on all mentioned above it is clear that paradox is a phenomenon that must be dealt with in a multidisciplinary way. Mathematicians, logicians and philosophers have been the most successful and are the most advanced in formulating the categories and classification of paradoxes.

**Types of Paradox**

When a certain paradox is classified in a certain classification group of paradoxes it is a paradox that can simultaneously be: a falsidical paradox, sorites paradox, or for example the limit paradox. Recognition of paradoxes, among other phenomena and what makes them paradoxes, is still not defined enough.

Willard Quine makes the distinction between three types of paradoxes. The same division is explained in the similar way by Stephen Francis Barker in his book *Philosophy of Mathematics*. The following classification was based on the examples of logical paradoxes and the paradoxes of mathematical logic: (W. V. Quine, 1966 [1962])

1. ***Veridical paradoxes (paradoxes of verification)***, are paradoxes the result of which is absurd, but presented in a way that makes it appear true. It is a situation seemingly impossible or contradictory but nevertheless true. <u>Examples</u>: *Frederic's birthday in the operetta Pirates of Penzance[7], Arrow's impossibility theorem, Simpson's paradox, the existence of an equal number of even and odd numbers according to Cantor's theory…*

2. ***Falsidical paradoxes*** are based on the results that do not just seem untrue but really are false. The proof derived by seemingly correct reasoning that leads to an illogical and false conclusion. <u>Examples</u>: various false proofs in algebra[8], *The horse paradox, the paradoxes of Zeno…* 3.***Antinomies*** are paradoxes that do not belong to the two previous types.[9] They are self-contradictory in the way they are presented or in reasoning, and the contradiction was caused by the adopted principles of reasoning. Seemingly correct reasoning proves that the reached conclusion is both true and false. <u>Examples</u>: *Greling-Nelson's paradox*, *Russell' paradox*, *Epimenides' paradox*.

---

[7] In the operetta The Pirates of Penzance by W.S. Dilbert and Sir Arthur Sullivan, a young man by the name of Frederic joins the pirates at the age of 21. The conditions of his contract prevent him from joining prior to his 21st birthday. Frederic was born on 29 February. He was 21 years old but he only celebrated his 5th birthday.

[8] The 2=1 proof. Augustus De Morgan version: assume that X=1, multiply both sides by X => $X^2$=X. Subtract 1 from both sides => $X^2$-1=X-1. Transform the left side as follows (X-1)(X+1)=X-1. Divide both sides by X-1 => X+1=1. When X is replaced by 1 in accordance with the original assumption we have that 2=1. *Note*: The error that leads to contradiction in this paradox is the division by X-1, assuming that X=1. The equation is divided by zero, which creates infinity on both sides of the equation because a finite value divided by zero results in infinity.

The difference between veridical and falsidical paradoxes is clear. What is not clear is the boundary between them based on the prohibition to have paradoxes that are somewhere in between thus classified logical opinions.

Quine pointed out that determining the type of a paradox could lead to disagreement, because:

> *"One man's antinomy can be another man's veridical paradox, and one man's veridical paradox can be another man's platitude."* (Quine, 1966 [1962])

Paradox is deeply connected to the basic adopted principles that differ from man to man and from group to group. In the attempt to form a general concept of paradoxes one may fall victim to subjectivism. One may make relative the basis of fact, of what must be recognized as indisputable. Herein lies the duality of the paradox phenomenon. Paradox is simultaneously the cause of doubt and the foundation of truth.

---

[9] Certain Serbian authors do not distinguish between paradoxes and antinomies. Rozalia Madarasz-Szilagyi in *Mathematical logic and set theory*.

Quine makes the analogy between "mathematical and physical myth"[10], although he does not consider the formed contradictions in physics as "confirmed" since mathematics is a "normative discipline" much more strictly formal than physics.

Mark Sainsbury classifies paradoxes according to how dependant from their "reality camouflage" the formulation is; what the relevance is of the paradox formulated in a way recognizable to the subject, which interprets a certain phenomenon. He classifies them and numbers them on the scale of 1 to 10, where 1 denotes paradoxes with a low "reality camouflage" (the barber paradox), and 10 denotes paradoxes the interpretation of which rocks the very foundations of thought (liar's paradox).

Sainsbury noticed that paradoxes can be grouped according to the subject. Thus we have paradoxes dealing with: space, time and infinity (like the paradoxes of Zeno) or indications of rational action (Newcomb's paradox, prisoner's dilemma paradox).

One of the basic definitions of paradox, according to their subject, divides them into:[11]

- *Ontological paradoxes*, dealing with real phenomena, in a real physical environment. They can be otherwise called *real paradoxes*.
- *Semantic paradoxes*, dealing with the explanation of real phenomena, which can be of theoretical (explanatory) character or observational character. They can be otherwise called *imaginary paradoxes*.

---

[10] Quine, W. V."On what there is".
[11] A similar division was made by Frank Ramsey, in his work The Foundations of Mathematics, 1925. He makes the distinction between:
    1. **Logics-mathematical** (Russell's paradox, Burali-Forti paradox ...)
    2. **Semantic**, also known as **epistemological** (Epimenides' paradox, Berry's paradox...)

In another form this division can be connected to Sorensen's view of the differentiation of paradoxes. In Roy Sorensen's aforementioned book there is a clearly defined division to:

1.*Scientific paradoxes*, created by a scientific understanding of the world, in which observation, experiment, calculation and measurement play a key role. It confirms that many of the scientific paradoxes have already been solved.

2.*Philosophical paradoxes*, originate in that part of the human desire for knowledge that still cannot be "profitably" incorporated into science. These paradoxes are outside the fields that define knowledge by scientific methods.

Valid literature dealing with paradoxes abounds with different types of paradoxes, as forms of grouping. But there is no desire to classify paradoxes. These types of paradoxes will be enumerated in this work, together with those paradoxes that I believe should be differentiated from the recognizable descriptions and I was not able to find any such definitions of them:

1. Vagueness paradoxes,
2. Self-reference paradoxes,
3. Limit paradoxes,
4. Mathematical paradoxes of physics,
5. Infinity paradoxes,
6. Visual paradoxes,

7.Mixed paradigms paradoxes.

**1. Vagueness paradoxes** are frequent paradoxes, created by a form of misunderstanding, caused by vague definitions. The vagueness can refer to:

- Real object (much – little, yellow – red,);
- Value function (good – bad, smart – stupid,).

The paradoxes caused by a vague boundary between two characteristics, caused by the possibility of relativity of terms, like: tall – short, long – short, strong – weak, or in the colour spectrum when one colour turns into another through a variety of tones.

*Sorites paradoxes* are special cases of *vagueness paradoxes*. They are caused by only one key word with a vague meaning. What does it mean to have a lot of money? To have 100.000 euro or 200.000 euro? Having 100.000 euro in Bangladesh does not mean the same as having them in New York.[12] The quantity has an importance in both cases. But, what does it mean to have a lot of money, or where lies the quantitative boundary between a lot and a little? Often, for businessmen, both amounts of money can mean very little. Every quantitative possibility of comparison, with no initially substantiated clearly defined differences (also manmade)[13], creates the possibility of *sorites paradox*.

Considered etymologically, the name for this type of paradox is derived from ancient Greek word Soros, meaning: multitude, pile, crowd. All the paradoxes dealing with quantity belong to the group of *Sorites paradoxes*. A frequently mentioned example of this type of paradoxes is a pile of sand. Removing a grain of sand at a time from the pile, a question arises. *When can we say that the sand is no longer a pile?*

---

[12] It does not have the same value to different classes in New York.
[13] National borders are also fictional.

**2. Self-reference** paradoxes are said to be the paradoxes in which:

*"A key term is defined by a totality it belongs to and all paradoxes contain a cycle of argumentation"*[14]

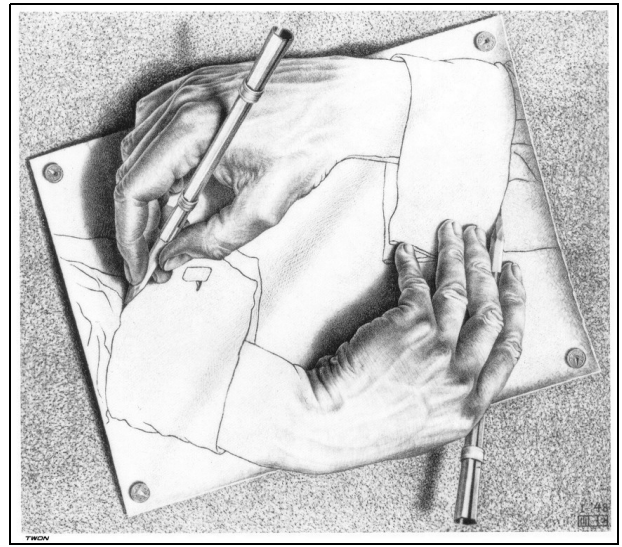

**Figure 1 M.C.Escher Drawing hands 17**

Self-reference paradoxes are caused by a special type of cyclic sophism that causes the paradoxical situation by "calling upon itself". The problem with this type of paradox is the difficulty to detect the self-reference. It is characteristic of these paradoxes that if split into constituents they contain no paradoxes. However, when whole they cause inconsistency in inference.

The appearance of these paradoxes is consistent with Gödel's incompleteness theorems that prove the possibility to create contradictions in formalized systems. Self-reference is the moment in which contradiction is realized. Self-reference appears in a cyclical connection of question and answer leading to a moment in which the answer is no longer consistent to the answer to the initial question of the cyclical progression.[15] Examples: *Epimenides paradox, Russell's paradox, Greling-Nelson paradox...*

---
[14] R. Madarasz –Szilagyi.

**3. Limit paradoxes** are, according *to Encyclopaedia Britannica,* the paradoxes that result from the wrong idea *that the limit (approximated) configuration created by, for example, reduction, has to exhibit the properties defined by the occurrence of corresponding properties of the approximated configurations*. In Quine's terms these are falsidical paradoxes, since an apparently correct reasoning leads to the wrong conclusion. Paradoxes of this type are mathematical in essence and are firmly rooted in physics. Paradoxical as it may initially seem, the final facts are arrived at by an endless series of steps, so that some of these paradoxes can be understood as infinity paradoxes.

A classical example of this type of paradox is the comparison of side lengths of an equilateral triangle ABC:

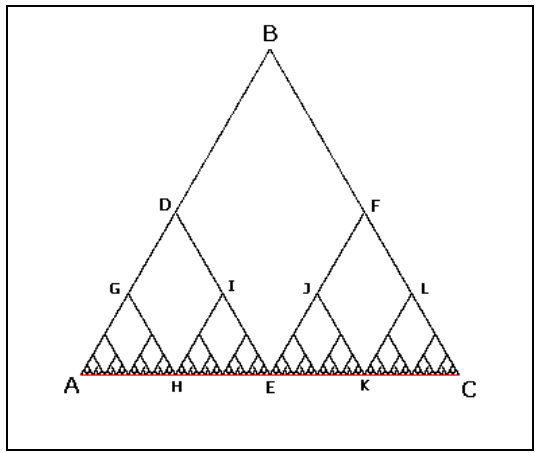

ABC=2AC
ADEFC=2AC
AGHIEJKLC=2AC
.
.
.
AC=2AC

Figure 2 Triangle division

---

[15] 1948 lithography

In physics these are the paradoxes that occur in borderline situations in which the nature itself provides limitations. These are the limits defined by physical constants, light speed, absolute zero, quantum of action, the limit of Schwarzwild surface. Examples: *D'alembert's paradox*, *STR paradoxes*, *Astronaut at the collapsing star event horizon paradox...*

**4. Mathematical paradoxes of physics** are paradoxes in which mathematical predictions of event development contradict real events. Like the limit paradox that could also be classified as a paradox of mathematical nature, this too is a *falsidical paradox* in Quine's terms. The paradox originates from the use of an apparently suitable mathematical system that leads to an obviously wrong conclusion. Examples: *D'alembert's paradox*, *Sommerfeld paradox*, *Gibbs paradox*, *Loschmidt paradox...*

**5. Infinity paradoxes** are classified as mathematical paradoxes, since mathematics has done the most, as the strictest analytical discipline of intellect, to shape the ideas of the meaning of infinity and its manifestations in abstract and real systems alike. Infinity denotes structures and processes without defined limitations and as such is very frequent in physics in the form of limit paradoxes. Examples: *Burali-Forti paradox*, *Galileo's paradox*, *Boltzmann's infinity paradox*, *Hilbert's paradox of the Grand Hotel*, *Monty Hall paradox*, *Shore and Mermaid paradox*, *Skolem's paradox*, *Poincare's paradox...*

**6. Visual paradoxes** are paradoxes that create visual illusion that clearly does not exist in reality. This type of paradox includes the triangle of Roger Penrose. The "wizard" of this type of paradox was Dutch painter and graphic artist Maurits Cornelius Escher (1902-1972). The interpretation of visual paradoxes is in direct connection with the causes of generated illusion. These paradoxes need no specially challenging explanation. Visual paradox can also be physical phenomena. There are real and imaginary visual paradoxes. The real ones are clearly nonexistent in the way we perceive them in reality (like a mirage) while the imaginary ones are fiction realized as a visual experience (like Escher's paintings or the triangle of Roger Penrose). These paradoxes are falsidical.
<u>Examples:</u> mirage, illusion that an object partially submerged in water is broken (spoon in a glass of water), *Penrose's triangle*...

**7. Mixed paradigms paradoxes** are paradoxes that appear when problems or phenomena are defined using conceptually different models of reasoning. The definition of this type of paradox is based on the views of Thomas Kuhn that conceptual changes of the "world image" occur as a consequence of the changes in the basic paradigms that generate it.

What is called a paradox when interpreting a phenomenon according to one paradigmatic pattern, does not have to be a paradox if the phenomenon is interpreted according to another paradigmatic pattern. Specifically, if the quantum-physical interpretation that a photon "simultaneously passes through both slits" agrees with the quantum-physical principles, that same interpretation is unacceptable to classical physics since it clashes with the principles that classical physics is based upon.

An example of how this paradox comes to be is when one paradigm of reasoning is mixed with another resulting in contradiction. If a certain problem is viewed from the standpoint of two basically different conceptual models, generalization produces contradictions that are called paradoxes. Another classical example of this type of paradox in physics is *Maxwell's demon*. When we introduce, in a context of microstates, described in terms of statistical physics, a "classical being" that behaves according to the laws of macrophysics, we arrive at the possibility of a paradoxical situation from the standpoint of a human observer.

Thomas Kuhn did not describe this type of paradox. They are obvious and recognizable and terminologically should be accepted as a distinct group of paradoxes. <u>Examples:</u> *Double slit paradox, Maxwell's demon paradox ...*

Paradoxes, regardless of the discipline of human spirit they belong to, can be solved, unsolved or "allegedly solved paradoxes". The last group consists of the paradoxes that are believed to be solved but the solution is incorrect. If paradoxes are confirmed as solved they can be called **exparadoxes.**

**Paradoxes in physics**

Paradoxes most often appear in "critical situations", when a new paradigmatic view in physics is formed; in a "crisis" of standardized opinion, that occurred due to the necessity of its modification caused by novelties that require corrections.

There are many different ways to classify paradoxes in physics. One of the ways, found on the Wikipedia (internet encyclopaedia)[16] site is this:

---

[16] The author of which is unknown to me.

1. ***Paradoxes relating to false assumption**s* (*Twins paradox*, *Ladder paradox*, *Supplee's paradox*, *Babinet's paradox*, *Mpemba paradox*, *Gibbs' paradox*, *Olbers' paradox*). These are the paradoxes that defy common sense predictions. The assumptions leading to the predictions were false or incomplete.

2. ***Paradoxes relating to unphysical mathematical idealizations*** (*Zeno's paradoxes*, *D'alembert's paradox*, *STR paradoxes*). This category also includes limit paradoxes caused by idealization that occurs when values are reached that are defined by physical constants.

3. ***Quantum mechanical paradoxes*** (*EPR paradox*, *Schrödinger's cat*, *Black hole information paradox*). This type of paradox uses the classification principle of division of physics into branches, which is totally unacceptable in this context.

4. ***Causality paradoxes*** (*Time travel paradox*, Loschmidt's paradox). Causality paradoxes are the paradoxes that violate the linear nature of causality that constitutes one of the primary principles of classical physics.

5. ***Observational paradoxes*** (*GZK paradox*). There are several kinds of observational paradoxes. Observational paradoxes also include the aforementioned visual paradoxes, although in this case, if we are talking gradation, the lowest form of paradox. The paradox I would especially like to mention can be found in Heisenberg's autobiography. In it he cites Einstein's opinion expressed in one of their discussions. Albert Einstein pointed out how observation is a very complex process. The event observed causes certain effects in the measurement device, which in turn form a sensory impression initiating a certain intellectual reaction of the subject. That is a lasting progression and goes through many successions that should be well understood in order to avoid errors. According to Einstein, theory is what leads along that progression and points to what should be really observed, and if there are certain contradictions and inconsistencies that cause paradoxes, it is because:

> *"although we are about to formulate new natural laws that do not agree with the old ones, we nevertheless assume that the existing laws—covering the whole path from the phenomenon to our consciousness—function in such a way that we can rely upon them and hence speak of observation."[17]*

The same source, *Wikipedia*, claims that paradoxes in physics are in apparent contradiction to physical description. Also, the difference is made between paradoxes accepted as a curiosity in physics (in majority) and paradoxes caused by inappropriate interpretation of theory (in minority). Generally speaking, it is concluded that paradoxes in physics and science in general suggest an error of some kind of incompleteness.

The above classification has a great flaw. There is no common reference according to which the classification was made, that can be said to be the distinction between paradoxes. In order to call something a classification it is necessary to determine a common denominator to all the classes of the classification.[18] The example is that quantum mechanics is a branch of physics and that the paradoxes caused by false assumptions can be quantum mechanical but that they can also belong to any other branch of physics. This means that these two differences cannot be part the same classification of paradoxes in physics.

---

[17] Heisenberg, W. (1969).
[18] Coen, M. & Najgel, E. (1934).

Aharov and Rohrlich (Aharov and Rohrlich, 2003) adopted Sainsbury's definition of paradoxes, but interpret it in a significantly more radical way. To them a paradox is very useful, since it points to the fact that something is wrong when everything seams to be right. ***According to them paradoxes point out that something is wrong, but not to what is wrong***. They classify paradoxes in physics in three classes according to the cause of the error:

1.**Paradoxes of error** (errors) caused by an error that can be simple or very subtle and difficult to recognize. If the error is discovered the paradox is solved. ***The basic characteristic of this paradox is that to solve it means to improve the understanding of theory, but not to improve the theory itself.*** An example is how erroneous understanding of synchrony in STR causes paradox. Examples: *Einstein-Bohr box paradox*, *Twins paradox*, *Dingle's paradox*, *Andromeda paradox*...

2.**Paradoxes of gaps** (gaps) also known as *paradoxes of contradiction* ***point out flaws in physical theory***. In the case of paradoxes of gaps the flaw is not of "fatal nature". Solution of the paradox does not interfere with the physical theory. Examples: *Wheeler's black hole entropy paradox*...

3.**Paradoxes of contradictions** (contradictions) are a more extreme version of paradoxes of gaps. ***This type of paradox points to a contradiction in physical theory that is of "fatal nature"***. Their solution changes the paradigm. Examples: *Paradox of preservation of charge in a rotating system*, *Ultraviolet catastrophe paradox*…

As the previous text shows there is a value gradation of paradoxes and there are types of paradoxes. However, all this relates to paradoxes of philosophy, mathematics and logic. How much of these formulations can be adopted in physics? Are there any formulations that have not been mentioned in previous systematizations?

The criteria for paradox classifications in physics can be divided according to:

I. the fields in which the paradoxes appear;

II. the form of formulation (whether they originate from empirical or theoretical subject of physics);

III. the degree of development of the fields within which they appear (in cases of formulating a new field or crisis of reasoning in a defined field).

I.

As a classification criterion paradoxes can be grouped according to the branch of physics they appear in. There are various divisions but since we must choose one, we choose this one:[19]

-Classical physics;

-Quantum physics;

-Relativity physics.

---

[19] Classical physics is physics that does not interpret phenomena; whose speed is approximate to the speed of light (special theory of relativity), with mass values in which gravity or acceleration are extremely large (general theory of relativity), in the areas of extremely small that is proportional to man as a subject that observes nature (quantum physics).

Ivan V. Anicin (Anicin, 2006) put forward the definition that coincides with my personal understanding of physics: "*Physics is describing nature by numbers.*"[20] This definition is necessary in order to see what destabilizes this "house" of numbers, what it is that makes it paradoxical in certain situations. Physics is an empirical science, the consistency of which, except in factual definition, is based on mathematics, the strictest normative discipline of the human mind. This made the distinction in physics between theoretical and experimental subjects, inseparable and complementary, when what is now known as physics was formalized.

II.

Paradoxes can be divided according to whether they originate in observational or normative mathematical theoretical analysis in physics:

1. ***Experimental paradoxes*** appear when laboratory experiments are performed the results of which have no satisfactory explanation;

2. ***Theoretical paradoxes*** appear when a theoretical explanation of a real physical phenomenon has not been precisely formulated. They are not definitely founded on real physical events. They are a significantly more frequent type of paradox and are caused by:

   a. different theoretical speculations with mathematical justification,

   b. mathematical speculations that cannot explain real physical events.

---

[20] Mathematics is the language of physics, in both theoretical and experimental aspect. Natural phenomena are quantified by numbers, which enables the correctness of their interpretation to be verified.

When *theoretical paradoxes* are specially considered it can be noticed that paradoxes are frequently presented by thought experiments, but that there are also paradoxes that originate directly from the explanation of physical phenomena. If the basis of theoretical speculation is used as a classification criterion, paradoxes can be:

1. Paradoxes that are thought experiments – paradoxes created as a result of theoretical speculation (*Twins paradox, Maxwell's demon paradox...*);

2. Paradoxes that are not thought experiments – paradoxes created by explaining physical phenomena that are sensory objective to man. (*Denny's paradox, Mpemba effect...*)

III.

Paradoxes in physics can differ in relevance, depending on how accepted they are and how adopted their physical explanation is by physical public.[21] Thus, paradoxes can be created:

1. as results within unconfirmed theories (*Black hole information paradox*);

2. as results within accepted theories (*relativistic paradoxes, Olbers' paradox*);

3. as results within the scope of the knowledge about physical processes (*Mpemba effect*).

Another possible basic division of paradoxes in physics is according to how defined the problem is and how possible the explanation is:

---

[21] If a certain explanation is more present in physical public, the paradox that appears within the explanation is more important. The importance quantitatively rises in proportion to the intention to correct the explanation until it is changed.

1.***Paradox with an explanation***. All physical paradoxes whose analysis eliminated the paradoxicality. I also call this type of paradox an exparadox (*Heat death paradox, Ultraviolet catastrophe paradox...*);

2.***Paradox without an explanation***. Paradox cases in which, in real physical systems, certain basically unexpected changes appear that lack a suitable and satisfactory explanation (*Ehrenfest paradox, GZK paradox...*)

From everything that has been said to explain paradoxes in physics it is obvious that they are based on different premises. The expressiveness of paradoxes in physics is fairly diverse. According to how a paradox in physics is formed and the reasons why a physical phenomenon is called a paradox at all, I propose this division of paradox types:

1.***pseudo paradox*** is not a paradox in the real sense of the word. If the phenomenon, at first believed contradictory, is precisely considered the contradiction is proved nonexistent. There is no element of the phenomenon that is confusing – contradictory to previous knowledge. Milorad Mladenovic calls this type of paradox an *alleged paradox*.[22] He assumes that there are "paradoxes" that do not deserve to be called that but he does not go into explanations or examples. (all the types of visual paradoxes, *Aristotle's wheel paradox, Hydrostatic paradox...*)

2.***paradox of idealization*** is created when a certain physical process is idealized. Occasionally, when a physical phenomenon is explained it needs to be idealized so that it looses its foundation in reality – it does not really exist. Idealization produces an approximated phenomenon described by fewer parameters than the real one. This formulates an approximated view that may or may not be similar to the operational results of the real event. (when a situation in ideal fluid is considered: *D'alembert's paradox, Sommerfeld paradox*, ... or for example the case of *Maxwell's demon* when the probability of a real transition of a molecule from one vessel to the other is extremely small...)

---
[22] Mladenovic M., Razvoj Fizike (elektromagnetizam), IRO "Gradevinska knjiga", Beograd, p. 122.

3.**hierarchical paradox** is a paradox created when a matter undergoes a state change. With this type of paradox there is no clear explanation of the reasons why a certain type of ontological principle (upon which the form of the existence of matter is based) is changed when changing to other physical states, clearly hierarchically differentiated. The paradox is formed by manifest difference of physical behaviour of matter: micro and macro state, state of aggregation or a theoretically assumed matter state in black holes. (*Boltzmann's paradox, Mpemba effect, Black hole information paradox*...)

4.**transitions paradox** is created in the process of developing the solution to a physical problem, when a theoretical or physically real phenomenon is explained. This type of paradox is inspirational and its basic tendency is to formulate an idea into a clearer and more consistent system. It is formed in the process of defining the explanation of a physical phenomenon when the theory has not yet been fully formulated. (*Wheeler's paradox, Loschmidt' paradox, Zermelo's paradox*...)

5.**paradox of assumption**, is a paradox in case of which the assumption explaining the phenomenon is itself intuitively not plausible enough (contrary to everyday experience), so the deduction based upon certain theoretical principles leads to contradiction with the expected real state of the physical system. There is a paradox within the theory which is consistent. However, the results of measurement cannot be proved or at least not proved enough. (*STR "greater" to "smaller" contradiction paradoxes, submarine paradox, twins' paradox, GZK paradox...*)

6.***paradox of paradigm*** is a paradox that exists within a paradigmatic point of view and when the paradigmatic point of view is changed the paradox disappears. If real experience contradicts theory, the change of important principle changes the explanation which is no longer contradictory to observation data. (*Ultraviolet catastrophe paradox, Olbers' paradox, Clausius' paradox, De Broglie's box paradox…*)

The question of continuum and discontinuum is of principal importance. The physical image of the world is built upon the basic position about the structure of nature. Since this is a very important difference we will define physical paradoxes based on this difference:

1. Paradoxes of continuum physics;
2. Paradoxes of discretion physics.

This division should distinguish between paradoxes belonging to continuum physics from those belonging to discontinuum physics. A good example of a continuum paradox is *ultraviolet catastrophe paradox*. Considering the problem of ultraviolet radiation from the continuous point of view the paradox is that the Earth should have been incinerated by the Sun's radiation. Considering the problem from the discontinuous point of view the paradox does not exist. This is also a *paradox of paradigm* since from one paradigmatic viewpoint the paradox exists and from the other it does not. In this case, however, one paradigmatic viewpoint does not exclude the other.

**Conclusion**

The very attempt to define paradoxes in physics is very challenging since it is necessary to define the border areas in physics that are not clearly drawn, which implies a multitude of inconsistencies. Also, what can be called a paradox, especially a paradox in physics, is another problem that has to be given argumentation, for none exists. Recognising defined criteria in the existing types of paradoxes is one of the basic tasks to be undertaken concerning paradoxes in physics, since they are piled up with no firm classification argumentation.

This work has presented some paradox classifications available in literature after which, by analyzing numerous paradoxes in physics, a classification was made according to shared recognizable characteristics in common to certain groups of paradoxes.

It has been noted that many examples that have been called paradoxes are proved not to be ones, when what could be called a paradox in physics is precisely defined. The classification has been formalized so that certain paradoxes could belong to more than one classification group. Note, for instance, the *Carroll's paradox* which is simultaneously a theoretical paradox and a paradox of assumption between which no clearly defined boundary exists.

The work leaves the possibility for many of its views to be more clearly expounded and as such provides a basis for further analysis of the problem: the existence of paradoxes in physics.